\documentclass[twocolumn]{article}

\usepackage{graphicx}% Include figure files
\usepackage{bm}% bold math

\def\a{\alpha}
\def\b{\beta}

\def\d{\delta}
\def\e{\varepsilon}

\def\r{\rho}

\def\D{\Delta}

\def\H{{\cal H}}

\def\dd{\mbox{d}}

\def\bra{\langle}
\def\ket{\rangle}
\def\av#1{\bra #1 \ket}
\def\ra{\rightarrow}

\def\nn{\nonumber}
\def\ni{\noindent}

\def\beq{\begin{equation}}
\def\eeq{\end{equation}}
\def\beqa{\begin{eqnarray}}
\def\eeqa{\end{eqnarray}}

\begin{document}

\twocolumn[
\begin{center}
{\Large \bf Direct derivation of microcanonical ensemble average from many-particle
quantum mechanics}\\
\bigskip

Tetsuro {\sc Saso}

{\it Department of Physics, Faculty of Sciences, Saitama University, Shimo-Ohkubo 255, Saitama-City, Saitama 338-8570 Japan.}
\medskip
\begin{quote}
Starting from the quantum mechanics for $N$ particles, we show that we can directly derive the microcanonical ensemble average of the physical quantity $A$ by using only the long time average and the equal probability assumption for the equal energy states.  The system is considered to be embedded in the outer world and we describe them in terms of the density matrix method.
\end{quote}
05.30-d \end{center}
\bigskip
]

\ni
{\it Introduction.---}
Foundation of the statistical mechanics is still an open problem.  It is often described in many textbooks of statistical mechanics ({\it e.g.}\cite{Greiner,Schwabl}) that one consider a gas of $N$ atoms and the motion of the point in a phase space of $6N$ dimensions consisting of $\{x,p\}\equiv\{\bm{x}_i,\bm{p}_i; i=1\sim N\}$.  The point moves on the equal energy surface in this phase space, and it is often assumed that the famous or notorious ``Ergod Theorem'' holds.  The probability that the state $\{x,p\}$ appears is assumed to be proportional to the volume of the phase space sandwitched by the equal energy surfaces with energy $E$ and $E+\D E$.  For quantum systems, this volume is quantized by the Planck constant $h$.  However, it is clear that such an old-fashoned treatment must be completely changed.  The quantum statistical mechanics must be directly derived from the quantum mechanics for $N$ particles with appropriate approximations.  We\cite{Saso} realize it by using the density matrix for the targeting system S plus the surrounding outer world (we denote it as B (Bath)).  Calculating the quantum mechanical average of the physical quantity $A$ and long-time average, we naturally obtain the form similar to the microcanonical ensemble average of $A$.  Application of the equal probability principle for the equal energy states, we reach the familiar form of $\av{A}$ in the microcanonical ensemble.

\medskip
\ni
{\it Time-Evolution and Time-Average of Isolated Many-Particle Systems.---}
\label{sec:isolated}
Consider first an isolated system including $N$ particles.  The wave function for $N$ particles is described by the wave function $\Psi(\bm{x}_1\cdots\bm{x}_N; t)$ which will be abbreviated as $|\Psi(t)\ket$.  {\it An observer stands outside the system, and the system remains isolated until he makes observation at a time $t$.}  He repeat the same observation with the same initial condition.  After many times of observations, the expectational value of a physical quantity $A$ becomes
\begin{equation}
  \av{A(t)} = \av{\Psi(t)|\hat{A}|\Psi(t)}.
\end{equation}
Generally, $|\Psi(t=0)\ket$ is described by the superposition of the set of eigenstates $\{|\Phi_n\ket\}$ and eigen energies $\{E_n$\}as
\begin{equation}
  |\Psi(t)\ket = \sum_n c_n e^{-iE_nt/\hbar} |\Phi_n\ket
  \label{eq:Psi}
\end{equation}
where $c_n$'s are the coefficients.  The expectational value of the total energy of the system is calculated  as
\begin{equation}
  \av{E}= \av{\Psi(t)|\hat{\H}|\Psi(t)} = \sum_n |c_n|^2 E_n,
\end{equation}
which does not depend on time.  But the expectational value of a quantity $A$ which does not commute with the Hamiltonian becomes
\\
\begin{equation}
  \av{A(t)} = \sum_{nm} c_n^* c_m e^{i(E_n-E_m)t/\hbar} \av{\Phi_n|\hat{A}|\Phi_m},
  \label{eq:A_t_N}
\end{equation}
{\it which oscillates in time for ever.  Namely, the system never reaches the thermal equilibrium.}
Consider a gas, for example.  If there is no mutual interaction between atoms of the gas, the wave function of the total system can be written by the product of the single particle wave function (here, the plane waves) $\phi_{\bm{k}}(\bm{x})$ as
\begin{equation}
  \Phi_n(x) = \prod_{i=1}^N \phi_{\bm{k}_i}(\bm{x}_i).
\end{equation}
(For simplicity, we do not consider (anti-)symmetrization of the wave function, which does not affect the following discussions.)
But if there is a weak mutual interaction, the atoms repeatedly collide with each other, and the distribution of $\{\bm{k}_i\}$ will gradually change.  The total wave function can not be described by the product of the plane waves but becomes the superposition of various product wave functions.   But can we still expect that the wave function will be relaxing  to proper one for the equilibrium state, which is similar to the product of the plane waves if the mutual interaction is weak?  Surely, most of the wave function $\Phi_n(x)$ will become an adequate one for the thermal equilibrium.  But in the quantum mechanics, the effect of the interaction is already included in the initial state $|\Psi(t=0)\ket$ and the time evolution of each eigenstate component is described by $e^{-iE_nt/\hbar}|\Phi_n(0)\ket$.  If a strange state $|\Phi_{\rm strange}(0)\ket$ was included in $|\Psi_n(t=0)\ket$ at the beginning, it remains for ever and never vanishes.  Therefore, {\it in the isolated systems, a relaxation to the thermal equilibrium does not occur}.  The wave function $|\Psi\ket$ remains to be the superposition of the eigenstates  as eq.(\ref{eq:Psi}).

But we like to know the averaged values of the physical properties in the thermal equilibrium, which do not depend on time.  Then, it is natural that what we observe may be the long time average of $\av{A(t)}$ over sufficiently long time $T$. 
{\it The time average may be done by integrating $\av{A(t)}$ in the range $[t-T/2,t+T/2]$} and divide by $T$, for example.  Then, we can write
$$  \overline{e^{i(E_n-E_m)t/\hbar}} = \frac{1}{T} \int_{t-1/2T}^{t+1/2T} e^{i(E_n-E_m)t'/\hbar} \dd t'$$
\begin{equation}
  = e^{i(E_n-E_m)t/\hbar}\frac{e^{i(E_n-E_m)T/2\hbar}-e^{-i(E_n-E_m)T/2\hbar}}{i(E_n-E_m)T/2\hbar} \label{eq:Joushiki}
\end{equation}
where over-line means the time average.  We set $E_n-E_m=\e$，$T/2\hbar=a$, and noting that $\av{A(t)}$ is real, we can write $\av{A(t)}=(\av{A(t)}+\av{A(t)}^*)/2={\rm Re}\av{A(t)}$.  Then, we obtain
\begin{equation}
  \mbox{eq.}(\ref{eq:Joushiki}) = \cos(\e t/\hbar)\frac{\sin a\e}{a\e}
  \label{eq:Joushiki2}
\end{equation}
$\sin (a \e)/a\e$ has a sharp peak at $\e=0$ as a function of $\e$ when $a=T/2\hbar$ is large enough, and otherwise quickly damps out with oscillation.  Furthermore, one can show that
\begin{equation}
  \int_{-\infty}^\infty \frac{\sin a \e}{a\e} \dd \e = \frac{\pi}{|a|}.
\end{equation}
Using the delta-function, we can regard as $\sin(a\e)/a\e \approx (\pi/|a|)\d(\e)$.  Thus, since $|\e|=|E_n-E_m|\ra 0$, we can regard as $\cos((E_n-E_m)t/\hbar)\ra 1$.  Then, we obtain
\begin{equation}
  \overline{e^{i(E_n-E_m)t/\hbar}}
  \stackrel{T\ra\infty}{\longrightarrow}   \frac{2\pi\hbar}{T}\d(E_n-E_m).
  \label{eq:time-average}
\end{equation}
Namely, only the equal energy states $E_n = E_m$ survive in $T\rightarrow \infty$, but note that not all the $E_n$'s are equal.  They are divided into many groups whithin which $E_n$'s are equal with each other. The energies are different for different groups.

When $E_n=E_m$, $e^{i(E_n-E_m)t/\hbar}=1$.  So, we can write as
\begin{equation}
  \overline{e^{i(E_n-E_m)t/\hbar}} \left.\longrightarrow  1 \right|_{E_n=E_m}.
  \label{eq:equi-energy}
\end{equation}
Thus, we can write as
\begin{equation}
  \overline{\av{A(t)}} = \sum_{nm}c_n^*c_m \left. \av{\Phi_n|\hat{A}|\Phi_m}\right|_{E_n = E_m}
\end{equation}
Here, $c_n^*c_m$ is a Hermitian matrix.   Then, we can diagonalize this matrix by a unitary transformation.  By making corresponding superposition of $\Phi_n$，$\Phi_m$, we obtain the eigenvectors $|\tilde{\Phi}_i\ket$ and eigenvalues $p_i=|\tilde{c}_i|^2 \ge 0$.  
Thus, we can write as
\\
\begin{equation}
  \overline{\av{A(t)}} = \sum_i p_i \av{\tilde{\Phi}_i|\hat{A}|\tilde{\Phi}_i}
  \label{eq:av-At}
\end{equation}
If we set $\hat{A}=1$, we obtain $\sum_i p_i=1$. Thus, {\it $p_i$ can be interpreted as a probability for the state $i$.}  By taking the time-average,  a statistical average is naturally introduced to an isolated system in addition to the quantum mechanical expectational values. Furthermore, the states are bundled into groups in each of which the energies are the same with each other  due to the time average.

Thus the conclusion in this section is that {\it a time-average of a measured value of a quantity $A$ in an isolated many-particle system can be expressed by the expectational value by the quantum mechanical wave function $\tilde{\Phi}_i$ plus the statistical average with the probability $p_i$, as in eq.(\ref{eq:av-At}).}

But in the truely isolated systems, any combination of $|\Phi_n\ket$ can be prepared in the initial state and they never change.  Therefore,  it is not approapriate to describe the thermal equilibrium states.  A solution to this problem will be presented in the next section.

\medskip
\ni
{\it Quantum Theory of Partial System and Density Matrix. ---}
\label{sec:DensityMatrix}
In the previous section, we investigate the isolated many particle systems quantum mechanically, and calculate the time average to measure the value of a quantity of the system.   In principle the isolated system can be a small sample of a cube of the volume 1cm$^3$, or a whole world.  But, we needed an observer outside the system, so that the system can not be a whole world.

\begin{figure}[h]
\centerline{\includegraphics[width=5cm]{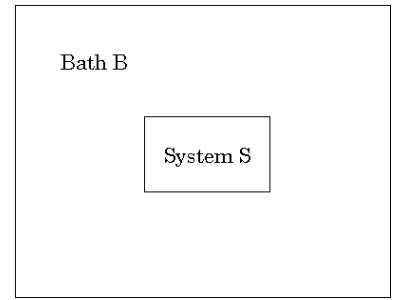}}
\caption{The world is divided to the partial system S and the bath B.}
\label{fig:world}
\end{figure}

In this section, we divide the world into two parts, one is the {\it partial system} (or simply the {\it system S}) which represents a sample to be measured, and the surrounding {\it outer world} or the {\it bath B} which represents the measurement apparatus like photo-plates.  Both the system S and the bath B constitute the world S+B (see Fig.\ref{fig:world}).  (The human beings can be included in the bath B in the same way as the measurement apparatus.)    In the followings, we will explain that if we look at only the partial system embedded in the world, we can regard the system S as a statistical ensemble of many equivalent quantum systems, in terms of the concept of the {\it density matrix}\cite{Feynman}.

Since the world is divided into the system S and the bath B, we also divide the $N$ variables ($\bm{x}_1, \cdots, \bm{x}_N$) in the wave function into those included in S and those in B.  The former is written as $x$, and the latter as $y$, for simplicity.  We assume that the particles can not transfer between S and B.  We allow weak interaction between the particles in S and B.  Then the Hamiltonian of the world may be written as
\begin{equation}
  \H_{\rm S+B}(x,y) = \H_{\rm S}(x) + \H_{\rm B}(y) + \H'(x,y)
\end{equation}
where the third term express a weak interaction between the particle in S and B.  We assume that there is no transfer of macroscopic size of energy or heat between S and B.
The wave function of the world can be written as $\Psi(x,y)$.

Using the complete orthonormal set $\{\phi_i(x)\} $, we can expand $\Psi(x,y)$ as a function of $x$ as
\begin{equation}
  \Psi(x,y) = \sum_i c_i(y)\phi_i(x).
\end{equation}
Here, $c_i(y)$ denotes the expansion coefficient, but it depends on $y$.  Using the orthonormal complete set $\{\psi_\a(y)\}$ in B, we can further expand as $y$ in the form
\begin{equation}
  c_i(y) = \sum_\a c_{i\a}\psi_\a(y),
\end{equation}
then, we obtain
\begin{equation}
  \Psi(x,y) = \sum_{i\a} c_{i\a} \phi_i(x) \psi_\a(y).
\end{equation}
Using the Dirac notation, we can write as
\begin{equation}
  |\Psi\ket = \sum_{i\a} c_{i\a} |\phi_i\ket |\psi_\a\ket.
\end{equation}

Now, suppose $\hat{A}_S$ is an operator acting only the variables in S, namely $x$, the expectational value of $\hat{A}_S$ becomes
\begin{eqnarray}
  \bra A_S\ket &=& \bra \Psi|\hat{A}_S|\Psi\ket = \sum_{ij\a\b}c_{i\a}^*c_{j\b} \bra \psi_\a|\bra \phi_i|\hat{A}_S|\phi_j\ket |\psi_\b\ket \nn \\
   &=& \sum_{ij\a}c_{i\a}^*c_{j\a} \bra \phi_i|\hat{A}_S|\phi_j\ket = \sum_{ij} (\r_S)_{ji} \bra \phi_i|\hat{A}_S|\phi_j\ket, \nn \\
   & &
   \label{eq:A}
\end{eqnarray}
where we used the property $\av{\psi_\a|\psi_\b}=\d_{\a\b}$, and define the $ji$ matrix element $(\r_S)_{ji}$ of the {\it density matrix}\index{density matrix} of the system S by
\begin{equation}
  (\r_S)_{ji} = \sum_\a c_{j\a}c_{i\a}^*.
  \label{eq:dmat}
\end{equation}
Also, we define the {\it density matrix operator} $\hat{\r}_S$ in such a way that the matrix elements of it by $\bra\phi_j|$ and $|\phi_i\ket$ coincide with $(\r_S)_{ji}$, namely, $(\r_S)_{ji} \equiv \bra \phi_j |\hat{\r}_S |\phi_i\ket$, then, eq.(\ref{eq:A}) can be written as
\\
\begin{equation}
  \bra A_S\ket = \mbox{Tr}_S (\hat{\r}_S \hat{A}_S).
\end{equation}
Tr denotes the ``trace'' in the matrix representation，and Tr$_{\rm S}$ means to sum over the diagonal indices regarding the system S.  $\hat{\r}_S$ is also called as the statistical operator of the system S.\index{statistical operator}

We define the {\it density matrix for S+B}  by $\hat{\r}_{S+B}\equiv |\Psi\ket \bra \Psi |$.  Then the matrix elements are given by
\begin{equation}
  (\r_{S+B})_{j\b,i\a} = \bra \phi_j|\bra\psi_\b|\hat{\r}_{S+B}|\psi_\a\ket|\phi_i\ket = c_{j\b}c^*_{i\a}
\end{equation}
so that we can write as
\\
\begin{equation}
  \bra A_S \ket = \mbox{Tr}_{S+B} (\hat{\r}_{S+B}\hat{A}_S)
\end{equation}
In addition, we can easily prove that {\it the density matrix of S is a trace over B of the density matrix of the world}：
\begin{equation}
  (\r_S)_{ji} = \sum_\a (\r_{S+B})_{j\a,i\a},
\end{equation}
namely,
\begin{equation}
  \hat{\r}_S = \mbox{Tr}_B (\hat{\r}_{S+B}).
  \label{eq:DensityMatrix}
\end{equation}

From eq.(\ref{eq:dmat}) we can show $((\r_S)_{ji})^*=(\r_S)_{ij}$.  Thus, $(\r_S)_{ji}$ is a Hermitian matrix, hence it can be transformed by a unitary transformation into a diagonal form with real diagonal matrix elements.  This is equivalent to the eigenvalue problem for the matrix $(\r_S)_{ji}$.  If we write the eigenvalues $p_n$ and the eigenvectors $|\tilde{\phi}_n\ket$, then we can write as
\begin{equation}
  \hat{\r}_S = \sum_n |\tilde{\phi}_n\ket p_n \bra \tilde{\phi}_n|.
\end{equation}
The matrix elements read
\begin{equation}
  (\r_S)_{ji} = \sum_n \bra \phi_j|\tilde{\phi}_n\ket p_n \bra \tilde{\phi}_n|\phi_i\ket.
  \label{eq:DensityMatrix2}
\end{equation}
As in the previous section, we can show that $p_n\ge 0$ and $\sum_n p_n=1$.

The expectational value of $A_S$ is written as
\begin{equation}
  \bra A_S \ket = \mbox{Tr}_S (\hat{\r}_S \hat{A}_S) = \sum_n p_n \bra \tilde{\phi}_n|\hat{A}_S| \tilde{\phi}_n\ket.
  \label{eq:av-A2}
\end{equation}
Here, $\bra \tilde{\phi}_n|\hat{A}_S| \tilde{\phi}_n\ket$ denotes the quantum mechanical expectational value of $\hat{A}_S$ in terms of $|\tilde{\phi}_n\ket$, and $\sum_n p_n$ denotes the statistical average with the probability $p_n$.  {\it In other words, this is a statistical (ensemble) average over the quantum state $|\tilde{\phi}_n\ket$ in S each of which appears with the probability $p_n$.
Namely, if the informations on the bath B are eliminated and confined into $p_n$, the expectational value of the quantity in S must be calculated not only by the quantum mechanical expectational value by the wave functions in S, but additionally by the statistical average with the probability $p_n$.
$\hat{\r}_S$ includes both effects.}

If $p_{n_0}=1$ for a single state $n_0$ and otherwise $p_n=0$, statistical operator becomes
\begin{equation}
  \hat{\r_S} = |\tilde{\phi}_{n_0} \ket \bra \tilde{\phi}_{n_0} |
\end{equation}
and the expectational value of $\hat{A}_S$ becomes
\begin{equation}
  \bra A_S \ket = \bra \tilde{\phi}_{n_0} |\hat{A}_S| \tilde{\phi}_{n_0} \ket
\end{equation}
which is the same form as the expectational value of the ordinary quantum mechanics.
This is called as the {\it pure state}\index{pure state}, and the others are called {\it mixed states}\index{mixed state}.
$\hat{\r}_S^2=\hat{\r}_S$ holds for a pure state.

\medskip
Thus, {\it if we divide the world into a small system S and the surrounding bath, and hiding the information on the latter into th density matrix, the quantum states of S becomes mixed states, which acquires a statistical character in addition to an ordinary quantum mechanical character.}
As a result, {\it partial systems can be regarded as a statistical ensemble of many quantum systems.}

\bigskip
This kind of discussion is already written in a few textbooks\cite{Feynman,Greiner}.  But thereby, $\{\tilde{\phi}_{n}\}$ are not assumed to be the eigenfunction of the system S.  Therefore, a time-evolution of the system and the average of the physical quantities have not been discussed.

We first consider the time-evolution of the system.  If the initial state was a superposition of the various eigenstates $\{\phi_i(x)\}$ of the system S, each   evolves as $\phi_i(x,t) = e^{-iE_it/\hbar}\phi_i(x)$, so that the density matrix (\ref{eq:dmat}) becomes
\begin{equation}
  (\r_S(t))_{ji} = \sum_{\a} c_{j\a}c_{i\a}^* e^{i(E_i-E_j)t/\hbar}.
\end{equation}
The expectational value of $A_S$ also evolves as
\begin{equation}
  \av{A_S(t)} = \sum_{ij} \sum_{\a} c_{j\a}c_{i\a}^* e^{i(E_i-E_j)t/\hbar}
  \av{\phi_i|\hat{A}_S|\phi_j}
\end{equation}

Here we regard the measurement of $A_S$ as taking the microscopically long-enough time  average as we have done in the previous section.  Then, we can replace
\begin{equation}
  \overline{e^{i(E_i-E_j)t/\hbar}}
  \stackrel{T\ra\infty}{\longrightarrow}  1|_{E_i=E_j}.
\end{equation}
Thus, we have
\begin{equation}
  \overline{A_S(t)} = \sum_{ij}\left. (\r_S)_{ij}\av{\phi_i|\hat{A}_S|\phi_j}\right|_{E_i=E_j}.
\end{equation}
Note again that it is usually the case that $\{E_i\}$ are divided into groups in each of which $\{E_i\}$ are equal with each other.
By taking the linear combination of $\phi_i$'s in each group, we can diagonalize the Hermitian matrix $(\r_S)_{ji}$ to obtain the eigenvalues $p_n$ and the eigen functions $\tilde{\phi}_n$,
then we can write as
\begin{equation}
  \overline{A_S(t)} = \sum_n p_n \av{\tilde{\phi}_n|\hat{A}_S|\tilde{\phi}_n}
\end{equation}
Since there is no matrix elements between the different groups, we can regard each group as independent.  If necessary, we can make a superposition of some groups.  Therefore, we consider only the group $E_n=E$:
\begin{equation}
  \overline{A_S(t)}(E) = \sum_n p_n \left.\av{\tilde{\phi}_n|\hat{A}_S|\tilde{\phi}_n}\right|_{E_n=E}
\end{equation}
{\it Thus, we have succeeded in deriving the formula in the same form as eq.(\ref{eq:av-At})  but all the energies $E_n$'s are equal with $p_n\ge 0$.}  The normalization is performed within the states with equal energy: $\left.\sum_n p_n\right|_{E_n=E}=1$.

This is the same form as eq.(\ref{eq:av-At}) for the isolated system, but any strange distribution is allowed in that case. Therefore, {\it the states that describe the thermal equilibrium can be derived only when we take the effect of the interaction with S and the outer world B correctly.}

Here we introduce the equal probability assumption $p_n={\rm const.}$ as usual.  Then, we reach the familiar expression for the expectational value of the quantity $A$ in the microcanonical ensemble.

\medskip
\ni
{\it Conclusions and Discussions ---}
We have derived the microcanonical ensemble average of the physical quantity $A$ in a form already well-known and described in most textbook in a simple and natural way, namely, by the long time average and the use of the density matrix for the partial system S.  We hope that all the textbooks for the statistical mechanics in the world may be rewritten following our approach.


\begin{thebibliography}{9}
\bibitem{Greiner} W. Greiner,L. Neize and H. St\"{o}ker: {\it Thermodynamics and Statistical mechanics} (Springer Verlag New York Inc., 1995)
\bibitem{Schwabl} F. Schwabl: {\it Statistical Mechanics} (Second Edition) (Springer Verlag, 2006)
\bibitem{Saso} T. Saso: {\it Statistical Mechanics} (in Japanese, Maruzen, 2010).  An English translation can be found in {\tt http://sces.th.phy.} {\tt saitama-u.ac.jp/$^\sim$saso/StatMech2013.pdf}.
\bibitem{Feynman} R. P. Feynman: {\it Statistical Mechanics} (Benjamin, 1972). 
\end{thebibliography}
\end{document}